\documentstyle[12pt]{article}

\def\ts{\thinspace}
\newcommand\Pop{\mbox{\rm P}}
\newcommand\Xop{\mbox{\boldmath X}}
\newcommand\Yop{\mbox{\boldmath Y}}

\newcommand\p{^\prime}

\newcommand\beq{\begin{equation}}
\newcommand\eeq{\end{equation}}
\newcommand{\rref}[1]{(\ref{#1})}
\newcommand\bear{\begin{array}}
\newcommand\enar{\end{array}}
\newcommand\Bear{\begin{eqnarray}}
\newcommand\Enar{\end{eqnarray}}
\newcommand\Bears{\begin{eqnarray*}}
\newcommand\Enars{\end{eqnarray*}}
\newcommand\nn{\nonumber}

\begin{document}
\baselineskip 1.2\baselineskip
\begin{flushleft}
{\it Yukawa Institute for Theoretical Physics}
\begin{flushright}
\begin{tabular}{l}
 YITP-95-4 \\
September 1995\\
hep-th/9509153
\end{tabular}
\end{flushright}
\end{flushleft}

\begin{center}
\vspace*{1.0cm}

{\LARGE{\bf   Non-linear Sigma Models on a Half Plane}}

\vskip 1.5cm

{\large {\bf M.\ts F.\ts  Mourad
\footnote{On leave of absence from Dept. Math.,  Minia Univ., Egypt.
 Partially supported by the Egyptian Missions Dept.}
 and R.\ts Sasaki
\footnote{Supported partially by the grant-in-aid for Scientific
Research,
Priority Area 231 ``Infinite Analysis'' and General Research (C) in Physics,
Japan Ministry of Education.}}}

\vskip 0.5 cm

{\sl Yukawa Institute for Theoretical Physics,} \\
{\sl Kyoto University,} \\
{\sl Kyoto 606-01, Japan.}
\vskip 0.5 cm

\end{center}

\vspace{1 cm}

\begin{abstract}
In the context of integrable field theory with boundary, the integrable
non-linear sigma models in two dimensions, for example, the $O(N)$,
the principal chiral, the ${\rm CP}^{N-1}$ and the complex Grassmannian
sigma models are discussed on a half plane.
In contrast to the well known cases of sine-Gordon, non-linear
Schr\"odinger and affine Toda field theories,
these non-linear sigma models in two dimensions are not classically
integrable if restricted on a half plane.
It is shown that the infinite set of non-local charges characterising the
integrability on the whole plane is not conserved for the free (Neumann)
boundary condition.
If we require that these non-local charges to be conserved,
then the solutions become trivial.
\end{abstract}

\vspace{1 cm}


\section{Introduction}
\setcounter{equation}{0}

Non-linear sigma models in $1+1$ dimensions or two dimensional
euclidean space have been discussed in various contexts in particle
physics:
as a theoretical laboratory for four dimensional gauge theories
\cite{ECS},
in the reduction of string theory, as a model field theory with
special geometrical features \cite{BCZ}, etc.
The non-linear sigma models in 2 ($1+1$) dimensions to be
discussed in this paper
are {\em harmonic maps} from a 2 dimensional space(-time) $M$ into
a target space $TG$.
In other words, they are ``free field theories"
and the Lagrangians consist of the kinetic terms only.
The non-linear structure of the target  manifolds is the origin of the
nontrivial and non-linear interactions.
It is well known that when the target space $TG$ is a group manifold or
a riemannian symmetric space \cite{Hel},
 the equation of motion can be expressed
in a Lax pair form and that the existence of an infinite set of
continuity equations is guaranteed \cite{ZM,BIZZ}.

The integrable structures and {\em classical solutions} of various non-linear
sigma
models have been investigated for more than a decade.
In particular, if the target space ($TG$) is a complex projective
space (${\rm CP}^{N-1}$) or complex Grassmannian manifolds,
very simple classes of solutions, the {\em instantons} and
{\em anti-instantons} \cite{WM}
are obtained by assuming the {\em finite action}.
They are  simply obtained from the two-dimensional
(anti) holomorphic functions, just like the ordinary two dimensional
harmonic functions are obtained as the real and/or imaginary parts of the
(anti) holomorphic functions.
In the language of harmonic maps they are called the
(anti) holomorphic maps.
Due to the conformal  invariance of the theory, the finite action
solutions can be considered as the solutions on the sphere
($M=S^2$), which is obtained from the two dimensional euclidean
space by the addition of the point at infinity (one point compactification).
Later a very general class of non-holomorphic solutions on $S^2$
for the $O(N)$ \cite{BG} and ${\rm CP}^{N-1}$ sigma models
\cite{DZa}\ts and the complex
Grassmannian sigma models \cite{RSa}\ts are obtained by a simple
algebraic method, the Gram-Schmidt orthonormalisation.
See also \cite{FKS} for the corresponding solutions of the
``supersymmetric" Grassmannian sigma models.

The exact and factorisable {\em quantum S-matrices} for various non-linear
sigma models have been known also for more than ten years \cite{BKZZ}.
They are obtained as solutions of Yang-Baxter equations with certain
prescribed symmetries.
In contrast the exact S-matrices of affine Toda field theories
\cite{AFZ,BCDSa,CM}
are diagonal and thus they are trivial solutions of
the Yang-Baxter equation.

\medskip
In the present paper we address the problem of the {\em non-linear
sigma models on a half plane. Can they retain the solvability
if some suitable boundary conditions are imposed?}

\medskip

   From the field theoretical point of view this problem can be considered
in the  context of
``integrable field theory on a half line" instead of  those on the whole line.
Here the central issue is whether the infinite set (or its suitable subset)
of {\em conserved quantities} in involution, which guarantees the
integrability,
can be preserved or not on the half plane.
Since the infinite set of {\em continuity equations} are a consequence of the
equation of
motion, which is independent of the boundary.
It is the {\em boundary conditions} that determine if the conserved quantities
are preserved or not.

For sine-Gordon  \cite{SkTa,GZa}, non-linear Schr\"odinger
\cite{SkTa}\ts and affine Toda field theories \cite{CDRS},
the integrable boundary conditions have been determined.
Thus for these theories, the main
objects of research is the effects of the boundary or the boundary
conditions which replace
the ``asymptotic conditions" in field theory on the whole line.
It should be remarked that in all the above mentioned theories
the {\em free boundary condition}
\footnote{In the linear theory it is called the Neumann boundary condition
${\partial u\over{\partial x_2}}=0$ at the boundary, $x_2=0$.},
i.e., zero boundary potential, is always integrable.

\medskip

 From the algebraic point of view this problem is interesting in connection
with certain extension of the Yang-Baxter equations and their related algebras.
Namely the non-linear sigma models on a half plane might offer
very interesting examples
of {\em factorisable scatterings with boundary} \cite{Cher}.
In this approach,
it is assumed that when a particle hits the boundary it is  reflected
elastically (up to  rearrangements among  mass degenerate particles).
The compatibility of the reflections and the scatterings constitutes the
main algebraic condition, called Reflection equation \cite{Cher,GZa,Goa},
which  extends the Yang-Baxter equation and the related algebras \cite{KSKS}.

\medskip

In this paper we concentrate on the aspects of the classical (or field
theoretical) integrability of the problem.
It is shown that the non-linear sigma model on a half plane with
{\em free boundary} fail to preserve the infinite set of non-local charges
which underlie the classical integrability on the whole plane.
In other words, the asymptotic conditions in field theory cannot always be
replaced
by boundary conditions at finite points
\footnote{Note that the periodic boundary conditions on a finite interval
always preserves the classical integrability.}
within the context of integrable field theory.

\medskip

This paper is organised as follows.
In section 2 we start with the two dimensional harmonic functions
which are a simplest
and well known example of harmonic maps.
Then the general two dimensional sigma models with
an arbitrary target space $TG$ is defined.
The boundary conditions and other necessary notions are introduced here.
In section three we briefly review the derivation of the infinite
set of  continuity equations after Br\'ezin et al. \cite{BIZZ}.
Section 4 gives the main result in the abstract form.
It is shown that for the non-linear sigma model with free boundary the
conservation of the infinite set of non-local charges is not satisfied
unless we additionally require that the first and second non-local
charges to be conserved.
This in turn result in an infinite set of conditions on the higher currents
at the boundary.
In sections 5, 6 and 7 we demonstrate the above result for the explicit models,
$O(N)$, the principal chiral and ${\rm CP}^{N-1}$ and the complex Grassmannian
sigma models.
It is shown that the requirement of the conservation of the second
non-local charge can only be met by {\em trivial solutions} among the
real analytic solutions.
Section 8 is devoted to summary and comments.

\section{Harmonic Maps}
\setcounter{equation}{0}

Let us start with the harmonic function on a half plane:
\Bear
-\infty <x_1<\infty \nn\\
0\leq x_2<\infty,
\label{hfplane}
\Enar
as the simplest example of the harmonic map
\footnote{Throughout this paper the two-dimensional space is euclidean,
$x=(x_1,x_2)$
and $x_1$ is considered as `euclidean time'.}.
It is described by the action
\beq
S=\int dx_1\left[\int_0^\infty {1\over2}(\partial_\mu u)^2dx_2 +
V_B(u)\right].
\label{harmfunc}
\eeq
In physics $u=u(x)$ is called a real scalar field in euclidean two dimensional
space.
The boundary potential $V_B$ is in general an arbitrary function of $u$
at the boundary, $x_2=0$.
For an integrable homogeneous boundary condition $V_B$ is quadratic in
$u$:
\beq
V_B(u)={a\over2}u^2.
\label{VBquad}
\eeq
Namely it attaches a `spring' with a spring constant $a$ at the boundary.

The action becomes stationary for the solutions of equation of motion
\beq
\triangle u=\partial_\mu^2u=(\partial_1^2+\partial_2^2)u=0,
\label{hareq}
\eeq
and the boundary condition
\beq
{\partial \over{\partial x_2}}u(x_1,0)=a\ts u(x_1,0),\quad {\rm mixed~b.c.}
\label{mixedbc}
\eeq
It is well known that the solution of the above problem is given by the real
and/or imaginary part
of  holomorphic functions (({\em anti}) {\em holomorphic map})
superposed appropriately in terms of the image charge method.
For the limits of $a\to\infty$ and $a=0$, the above boundary conditions become
the well known
\Bears
u(x_1,0)&=&0,\quad {\rm Dirichlet~b.c.}\quad a\to\infty, \\
{\partial \over{\partial x_2}}u(x_1,0)&=&0,\quad {\rm Neumann~b.c.} \quad a=0.
\Enars

\medskip

It is very easy to go from harmonic functions to a non-linear sigma model with
target space $TG$.
Different sigma models are obtained for different choices of $TG$.
Let
$$
u^\alpha,\quad \alpha=1,\ldots,N,
$$
be some local coordinates  of $TG$ with the metric tensor $g_{\alpha\beta}(u)$.
Then a non-linear  sigma model (or free field theory taking value in $TG$)
is defined by
\beq
S=\int dx_1\left[\int_0^\infty {1\over2}g_{\alpha\beta}(u)
\partial_\mu u^\alpha\partial_\mu u^\beta dx_2+ V_B(u)\right].
\label{gensigact}
\eeq
However, there is one big difference. That is the boundary term $V_B$.
In the harmonic function case, the quadratic boundary term \rref{VBquad}\ts has
a well defined meaning. On an arbitrary manifold, quadratic functions
of the local coordinates or other functions do not have an invariant meaning.
Therefore in this paper we consider only the {\em free boundary} case:
\beq
V_B=0.
\label{fbcond}
\eeq
The equation of motion is
\beq
\partial_\mu\left(g_{\alpha\beta}(u)\partial_\mu u^\beta(x)\right)=0,
\label{geneq}
\eeq
and the free boundary condition is
$$
g_{\alpha\beta}(u)\partial_2 u^\beta(x)=0,\quad {\rm at~the~boundary},
$$
or
\beq
\partial_2u^\alpha(x_1,0)=0.
\label{genfbc}
\eeq
In all the explicit examples treated below,
the Lagrangians have much simpler forms reflecting the high degrees of
symmetry of the systems.

\medskip

Before closing this section let us remark on the general setting of
harmonic maps of manifolds with boundary \cite{Ham}.
Let \Xop\ts and \Yop\ts be compact riemannian manifolds with boundary.
In \cite{Ham}\ts certain existence theorems of the harmonic maps
$f: \Xop\ts \to \Yop$ corresponding to the Dirichlet, Neumann and mixed
boundary conditions were given
\footnote{Here the boundary conditions were not derived from variations.}.
In all these cases the target manifold \Yop\ts was assumed to have
non-positive riemannian curvature, which was necessary for a heat
equation method to work.
These results seem to be irrelevant to the non-linear sigma models
in this paper, since they all have positive riemannian curvature.

\section{Non-Local Charges} 
\setcounter{equation}{0}

In this section we follow the argument of Br\'ezin et al \cite{BIZZ} (see
also \cite{CZa}),
and derive the infinite set of continuity equations.
Let us start by assuming that the `gauge' field
$$A_\mu(x)=A_\mu^{\alpha\beta}(x),\quad \alpha,~\beta=1,2,\ldots, N,\quad
\mu=1,2
$$
satisfy the following two properties for {\em the solutions of the
equation of motion},
\begin{enumerate}
\item[(i)]
$A_\mu$ is a `pure gauge', namely the corresponding `field strength' vanishes
\beq
F_{\mu\nu}=\partial_\mu A_\nu-\partial_\nu A_\mu +[A_\mu,A_\nu]=0.
\label{pgauge}
\eeq
\item[(ii)]
 $A_\mu$ satisfy the `continuity equation'
\beq
\partial_\mu A_\mu(x)=0.
\label{conteq}
\eeq
\end{enumerate}
In all three examples discussed below, the `pure gauge' condition is
trivially satisfied and the `continuity equation' has the dynamical
contents.
Whereas in the pseudodual chiral models \cite{ZM,CZa}, the `continuity
equation'
is satisfied trivially and the `pure gauge' condition has the non-trivial
dynamical meaning.

\medskip
In either case, the above two properties (i) and (ii) are encoded neatly
into the following
`linear scattering' problem
\begin{eqnarray}
	\partial_+\psi^\alpha & = &
	-\sum_\beta{A_+^{\alpha\beta}\over{1+\lambda}}\psi^\beta,
	\label{linp} \\
	\partial_-\psi^\alpha & = &
	-\sum_\beta{A_-^{\alpha\beta}\over{1-\lambda}}\psi^\beta,
	\label{linm}
\end{eqnarray}
in which $\lambda$ is the `spectral parameter' and $x_\pm$ are the
two-dimensional complex coordinates
\begin{equation}
	x_\pm=x_1\pm ix_2, \quad \partial_\pm={\partial\over{\partial x_\pm}},
	\label{compc}
\end{equation}
and
$$
    A_\pm={1\over2}(A_1\mp iA_2).
$$
In other words, the two conditions (i) and (ii)  can be  written as
`one parameter family of zero-curvature conditions'
\begin{equation}
	d{\cal A}-{\cal A}^2=0,\quad {\cal A}\equiv -{A_+\over{1+\lambda}}dx_+ -
	{A_-\over{1-\lambda}}dx_-,
	\label{zero}
\end{equation}
which represents the integrability of
 \rref{linp}\ts and \rref{linm}\ts expressed in terms of a
 `one parameter family of one form' ${\cal A}$,
\begin{equation}
 	d\psi={\cal A}\psi.
 	\label{onef}
\end{equation}
 Namely, the the two conditions (i) and (ii) which will be used below
 to derive the infinite set of `continuity equations' are equivalent with the
`Lax pair' (or the `linear scattering' problem) formulation of the
non-linear sigma models.

\medskip
If we define the `covariant derivative' operator
$$D_\mu^{\alpha\beta}=\delta^{\alpha\beta}\partial_\mu +A_\mu^{\alpha\beta}(x),
$$
then \rref{pgauge}\ts can be expressed as
\beq
[D_\mu, D_\nu]=0.
\label{zerocurv}
\eeq
Then \rref{conteq}\ts can be rewritten as an operator identity:
\beq
\partial_\mu D_\mu=D_\mu\partial_\mu.
\label{opident}
\eeq

Based on \rref{zerocurv}\ts and \rref{opident}, Br\'ezin et al
\cite{BIZZ}\ts showed the
existence of an infinite number of continuity equations inductively.
Suppose $J_\mu^{(n)}$ is the $n$-th conserved current,
$$
\partial_\mu J_\mu^{(n)}(x)=0,
$$
then there exists a function $\chi^{(n)}(x)$, such that
\beq
J_\mu^{(n)}(x)=\epsilon_{\mu\nu}\partial_\nu\chi^{(n)}(x),\quad n\ge1.
\label{poteq}
\eeq
Here $\epsilon_{\mu\nu}$ is the two-dimensional anti-symmetric tensor
with $\epsilon_{12}=1$. The next conserved current is defined
as
\beq
J_\mu^{(n+1)}(x)=D_\mu\chi^{(n)},\quad n\ge0.
\label{newcurrent}
\eeq
This definition is consistent if we identify
\beq
\chi^{(0)}(x)=1,\quad J_\mu^{(1)}(x)=A_\mu(x).
\label{firstcurr}
\eeq
It is easy to see that $J_\mu^{(n+1)}(x)$ satisfies the continuity equation
\Bear
\partial_\mu J_\mu^{(n+1)}(x)&=&\partial_\mu D_\mu\chi^{(n)}
=D_\mu\partial_\mu\chi^{(n)}
=-\epsilon_{\mu\nu}D_\mu J_\nu^{(n)}\nn\\
&=&-\epsilon_{\mu\nu}D_\mu D_\nu\chi^{(n-1)}=-[D_1,D_2]\chi^{(n-1)}=0.
\label{newcont}
\Enar

\setcounter{equation}{0}
\section{Conserved Quantities}

In the previous section we have seen that the
infinite set of continuity equations is a consequence of the equation of
motion.
But a continuity equation itself does not give a conserved quantity.
In order to obtain the conserved quantities from these
continuity equations
the boundary conditions play an essential role \cite{GZa,CDRS}.
Here we show that for the free boundary condition on the half plane
the infinite set of non-local charges are not conserved in general.
The conservation of these non-local charges imposes  very
severe constraints which eventually reduce the theory \lq trivial\rq.
We will not discuss the involution property of the corresponding non-local
conserved charges on the full plane.
(See the comments at the end of this section.)

As discussed in the section two we concentrate on  the free boundary,
$$
V_B=0.
$$
In this case  the non-local charges do not get any contributions from
the `boundary term' and are given simply by the integral of the currents.
Thus the first non-local charge is given by
\beq
I^{(1)}(x_1)=\int_0^\infty J_1^{(1)}(x_1,x_2)dx_2.
\label{1stcharge}
\eeq
Let us check if it is really conserved or not.
\beq
{d\over{dx_1}}I^{(1)}(x_1)=\int_0^\infty{\partial\over{\partial x_1}}
J_1^{(1)}dx_2
=-\int_0^\infty{\partial\over{\partial x_2}} J_2^{(1)}dx_2=J_2^{(1)}(x_1,0).
\label{bcontr}
\eeq
Here we have used the continuity equation \rref{conteq}\ts and assumed that
the currents vanish at $x_2=\infty$:
$$
\lim_{x_2\to\infty}J_1^{(1)}(x_1,x_2)=0=\lim_{x_2\to\infty}J_2^{(1)}(x_1,x_2).
$$
Therefore the conservation of the first charge $I^{(1)}(x_1)$
(i.e. independent of $x_1$)
is only achieved when the condition
\beq
J_2^{(1)}(x_1,0)=0
\label{cubc}
\eeq
is satisfied
\footnote{
As we see in the explicit examples in sections 5,6,7 this condition is
usually equivalent with the free boundary condition on the field.}.
Namely, the current flowing to the endpoint must be vanishing.

Likewise the $n$-th non-local charge is given by the integral of the
$n$-th current $J^{(n)}_\mu$,
$$
I^{(n)}(x_1)=\int_0^\infty J_1^{(n)}(x_1,x_2)dx_2.
$$
Since the above argument applies equally well to the $n$-th
current $J^{(n)}_\mu$, an additional  condition
\beq
J_2^{(n)}(x_1,0)=0
\label{ncubc}
\eeq
\medskip
must be met in order for it to be conserved.

Let us look into details. Let us start with the second current
$J_\mu^{(2)}$.
A scalar `potential' $\chi^{(1)}$ is defined by
\beq
J^{(1)}_1=\partial_2\chi^{(1)},\quad J^{(1)}_2=-\partial_1\chi^{(1)}.
\label{1stpot}
\eeq
Then the second current is defined by $J_\mu^{(2)}=D_\mu\chi^{(1)}$.
To be more specific
\Bear
J_1^{(2)}&=&\partial_1\chi^{(1)}+A_1\chi^{(1)}=-J_2^{(1)}+J_1^{(1)}\chi^{(1)},
\label{seccura}\\
J_2^{(2)}&=&\partial_2\chi^{(1)}+A_2\chi^{(1)}=J_1^{(1)}+J_2^{(1)}\chi^{(1)}.
\label{seccurb}
\Enar
Thus in order that $I^{(2)}$ is conserved
\beq
J_2^{(2)}(x_1,0)=J_1^{(1)}(x_1,0)+J_2^{(1)}(x_1,0)\chi^{(1)}(x_1,0)=0
\label{secbc}
\eeq
is necessary.
Since $A_2(x_1,0)=J_2^{(1)}(x_1,0)=0$, the above condition is equivalent to
\beq
J_1^{(1)}(x_1,0)=0=A_1(x_1,0).
\label{secbcc}
\eeq
 From  \rref{seccura}\ts this in turn implies
\beq
J_1^{(2)}(x_1,0)=0.
\label{addsecbc}
\eeq

\medskip
By induction we can show that if the first and second non-local charges are
conserved then the higher non-local charges $I^{(1)},\ldots,I^{(n)}$
are also conserved and that this also implies that both ($\mu=1,2$) components
of
the higher currents vanish at the boundary
\beq
J^{(k)}_1(x_1,0)=0=J^{(k)}_2(x_1,0),\quad k=1,\ldots,n.
\label{ncurbc}
\eeq

The next currents are expressed as
\Bears
J_1^{(n+1)}&=&\partial_1\chi^{(n)}+A_1\chi^{(n)}=
-J_2^{(n)}+A_1\chi^{(n)},\\
J_2^{(n+1)}&=&\partial_2\chi^{(n)}+A_2\chi^{(n)}=
J_1^{(n)}+A_2\chi^{(n)}.
\Enars
Thus by the assumption of the induction we find that the next charge is
also conserved and
\Bear
J_1^{(n+1)}(x_1,0)&=&-J_2^{(n)}(x_1,0)+A_1(x_1,0)\chi^{(n)}(x_1,0)=0,\nn\\
J_2^{(n+1)}(x_1,0)&=&J_1^{(n)}(x_1,0)+A_2(x_1,0)\chi^{(n)}(x_1,0)=0.
\label{ncurbcind}
\Enar
Therefore the conservation of non-local charges on a half plane
imposes very severe constraints.

\medskip
As a small digression, let us consider the case that $I^{(1)}$ and $I^{(3)}$
are
conserved but that the conservation of $I^{(2)}$ is not required.
In this case
\Bear
J_1^{(3)}&=&-J_2^{(2)}+A_1\chi^{(2)},
\label{3cura}\\
J_2^{(3)}&=&J_1^{(2)}+A_2\chi^{(2)}.
\label{3curb}
\Enar
Thus requiring $J_2^{(3)}(x_1,0)=0$ means
\beq
J_1^{(2)}(x_1,0)=0,
\label{twocura}
\eeq
which in turn (by \rref{seccura} ) means (by using $J_2^{(1)}(x_1,0)=0$ )
\beq
A_1(x_1,0)=J^{(1)}_1(x_1,0)=0
\label{twocurb}
\eeq
and $I^{(2)}$ is also conserved.

It should be remarked that  throughout the above discussion we used the fact
that
\beq
\chi^{(n)}(x_1,x_2)
\label{chibc}
\eeq
is non-singular at the boundary $x_2=0$.

\medskip

It is worthwhile to comment on the relationship with other approaches
to the conserved charges.
On the full plane there are many work on how to extract various sets of
infinite conserved charges from the `one parameter family of
zero-curvature' equations \rref{zero}, \cite{olds,CZa}, which roughly
correspond to  the various choices of the  expansion points,
$\lambda=\lambda_0$. Namely, the variety of infinite sets of conserved
charges can be understood as different explicit expressions of the `one
parameter family of conserved charges'.

\setcounter{equation}{0}
\section{$O(N)$ Sigma Model}

Next let us consider the consequences of the above severe constraints
coming from the
conservation of non-local charges for explicit field theories.
The first example is the well known $O(N)$ sigma model which consists of
an $N$-component real scalar field
$$
\phi=\{\phi^\alpha\},\quad \alpha=1,\ldots,N,
$$
satisfying the condition
\beq
\phi\cdot\phi=\sum_{\alpha=1}^N(\phi^\alpha)^2=1.
\label{oncond}
\eeq
Namely the target space is the $N-1$ dimensional unit sphere $TG=S^{N-1}$
in the $N$ dimensional euclidean space.
The model is defined by the action
\beq
S={1\over2}\int dx_1\int_0^\infty dx_2\sum_{\alpha=1}^N
\left(\partial_\mu\phi^\alpha\right)^2.
\label{onact}
\eeq
Since we mainly discuss the classical theory the coupling constant is
irrelevant. From the stationarity of the action we obtain
the equation of motion
\beq
\partial_\mu^2\phi+\phi(\partial_\mu\phi\cdot\partial_\mu\phi)=0,
\label{oneq}
\eeq
and the free boundary condition
\beq
\partial_2\phi(x_1,0)=0.
\label{onfbc}
\eeq
The first current is given by
\beq
J_\mu^{(1)}=A_\mu=A_\mu^{\alpha\beta}=2\left(\phi^\alpha\partial_\mu\phi^\beta
-(\partial_\mu\phi^\alpha)\phi^\beta\right).
\label{on1cur}
\eeq
Thus the conservation of the first charge is a consequence of the free
boundary condition.
It is straightforward to verify the continuity equation for $A_\mu$
\beq
\partial_\mu A_\mu^{\alpha\beta}=0,
\label{onconteq}
\eeq
by using the equation of motion \rref{oneq}.
It is also straightforward to show that the
current $A_\mu$ is a `pure gauge':
$$
\partial_1A_2-\partial_2A_1+[A_1,A_2]=0.
$$

\medskip
The conservation of the first and second charges is equivalent with
the condition
\beq
A_\mu^{\alpha\beta}(x_1,0)=0,\quad \mu=1,2.
\label{ononebc}
\eeq
By multiplying $\phi^\alpha$ we obtain
\beq
\partial_\mu\phi^\beta-(\phi^\alpha\cdot\partial_\mu\phi^\alpha)\phi^\beta=0.
\label{ononebcphi}
\eeq
Since $\phi\cdot\phi=1$ implies $\phi^\alpha\cdot\partial_\mu\phi^\alpha=0$ at
any point including the boundary, we obtain
$$
\partial_\mu\phi^\beta=0, \quad {\rm at~ the ~boundary},\quad\beta=1,\ldots,N,
$$
or
\beq
\partial_1\phi^\beta(x_1,0)=0=\partial_2\phi^\beta(x_1,0).
\label{ononebcphiex}
\eeq
The first equation implies
\beq
\partial_1^n\phi^\beta(x_1,0)=0,\quad n\ge1.
\label{onnbca}
\eeq
It is now obvious that the second charge is not conserved in general
and its conservation requires very strong conditions,
which allow only trivial solutions as we will see below.

Next let us calculate
$$
\partial_2^2\phi(x_1,0)
$$
by using the equation of motion at small $x_2$:
\Bear
\partial_2^2\phi(x_1,x_2)&=&\left(\partial_1^2+
\partial_2^2-\partial_1^2\right)\phi(x_1,x_2)
\nn\\
&=&-\phi(x_1,x_2)(\partial_\mu\phi\cdot\partial_\mu\phi)(x_1,x_2)-
\partial_1^2\phi(x_1,x_2).
\label{ontwobc}
\Enar
By taking the limit $x_2\to0$ in the above expression and using
\rref{ononebcphiex},
 \rref{onnbca}, we obtain
\beq
\partial_2^2\phi(x_1,0)=0.
\label{on2ybc}
\eeq
Similarly, by further differentiation of \rref{ontwobc}\ts we obtain
\Bear
\partial_2^3\phi(x_1,x_2)
&=&(\partial_2\phi(x_1,x_2))(\partial_\mu\phi\cdot\partial_\mu\phi)(x_1,x_2)\nn\\
&+&
\phi(x_1,x_2)\partial_2(\partial_\mu\phi\cdot\partial_\mu\phi)(x_1,x_2)
-\partial_1^2\partial_2\phi(x_1,x_2).
\label{onthrbc}
\Enar
Again by using \rref{ononebcphiex}\ts and
 \rref{onnbca}, we obtain
\beq
\partial_2^3\phi(x_1,0)=0.
\label{onthrbcvan}
\eeq
By repeating the same argument, we arrive at
\beq
\partial_2^n\phi(x_1,0)=0,\quad n\ge1,
\label{onnbcvan}
\eeq
which together with \rref{onnbca}\ts would exclude any non-trivial solutions
with
real analytic dependence on $x_2$.

\setcounter{equation}{0}
\section{Principal Chiral Model}

The next example is the principal chiral model, that is the sigma model
taking value in a group manifold.
It is defined by the action
\beq
S=\int dx_1\int_0^\infty dx_2{\rm Tr}\left(\partial_\mu g\partial_\mu
g^{-1}\right),
\label{prchirlag}
\eeq
in which $g=g(x)$ takes value in a group of $N\times N$ matrices.
The lowest member of the current is given by
\beq
A_\mu=g^{-1}\partial_\mu g,
\label{prchicur}
\eeq
which is obviously a pure gauge:
$$
\partial_1A_2-\partial_2A_1+[A_1,A_2]=0.
$$
The equation of motion reads
\beq
\partial_\mu A_\mu=\partial_\mu(g^{-1}\partial_\mu g)=0
\label{prchieq}
\eeq
and the free boundary condition is given by
\beq
(g^{-1}\partial_2g)(x_1,0)=0.
\label{chifbc}
\eeq
Therefore the conservation of the first charge is a consequence of
the free boundary condition, too.
The condition for conservation of the first and second charges
can be written succinctly
$$
g^{-1}\partial_\mu g=0,\quad {\rm at}\quad x_2=0
$$
or
\beq
\partial_\mu g(x_1,0)=0, \quad \mu=1,2.\label{prchivan}
\eeq
Again the conservation of the second charge is not guaranteed in general.

We will show below that the solutions of the equation of
motion preserving the infinite set of non-local charges are trivial.
 From the equation of motion, we obtain
\beq
\left(\partial_1^2g+\partial_2^2g\right)=(\partial_\mu g)g^{-1}(\partial_\mu
g).
\label{prchieqsec}
\eeq
By going to the boundary ($x_2\to0$) and by using \rref{prchivan}\ts we obtain
\beq
\partial_2^2g(x_1,0)=0.
\label{prchitwobc}
\eeq
And from \rref{prchivan}\ts we find also
\beq
\partial_1^ng(x_1,0)=0,\quad n\ge1.
\label{prchixnbc}
\eeq
By repeating similar arguments we can show
\beq
\partial_2^ng(x_1,0)=0,\quad n\ge1.
\label{prchiynbc}
\eeq
As in the case of the $O(N)$ sigma model
this would exclude any non-trivial solutions
with real analytic dependence on $x_2$.

\setcounter{equation}{0}
\section{${\rm CP}^{N-1}$ and Grassmannian Sigma Models}

There are many ways to express the ${\rm CP}^{N-1}$ and
Grassmannian sigma models.
The complex Grassmannian manifold $Gr(N,m)$ is a space of $m$-frames
in the complex $N$-dimensional vector space
$C^N$. Let $X$ be an $m$-frame (an $N\times m$ matrix)
$$
X=\left(e_{j_1},\ldots,e_{j_m}\right),
$$
in which $\{e_j\}$, $j=1,\ldots,N$ is an orthonormal basis of $C^N$:
$$
e_j^\dagger\cdot e_k=\delta_{jk}.
$$
Then $X$ satisfies the constraint
\beq
X^\dagger X=1_m,\quad m\times m ~{\rm unit ~matrix}.
\label{xdefs}
\eeq
We choose as a field variable of the $Gr(N,m)$ model
 a projector $\Pop=\Pop(x)$, which is an $N\times N$ matrix
\beq
\Pop=\Pop(x)=X(x)X(x)^\dagger,
\label{pdef}
\eeq
Then it is obvious that \Pop\ts has the properties of the projector:
$$
\Pop^2=\Pop,\quad \Pop^\dagger=\Pop.
$$
The special case of $m=1$ corresponds to the ${\rm CP}^{N-1}$ model.

The first current is defined by
\beq
A_\mu=2[\Pop,\partial_\mu\Pop],
\label{gracurdef}
\eeq
and the equation of motion reads
\beq
0=\partial_\mu A_\mu=2[\Pop,\partial_\mu^2\Pop].
\label{greq}
\eeq

They can be obtained by embedding (reduction) the $Gr(N,m)$ sigma model into
the
principal chiral model :
\beq
g=1-2\Pop.
\label{gremb}
\eeq
This has the properties
\Bears
g^2&=&(1-2\Pop)^2=1-4\Pop+4\Pop^2=1,\\
g^\dagger&=&1-2\Pop^\dagger=g.
\Enars
It is easy to derive \rref{gracurdef}\ts and \rref{greq}\ts from
\rref{prchicur}\ts and \rref{prchieq}, respectively.
The free boundary condition \rref{chifbc}\ts is rewritten as
\beq
\partial_2\Pop(x_1,0)=0.
\label{grfbc}
\eeq

As in the $O(N)$ and the principal chiral models the second charge is not
conserved in general, unless an additional condition
\beq
\partial_1\Pop(x_1,0)=0,\label{grsadcon}
\eeq
is satisfied.
If we require that the first and second charges to be conserved, then
the solutions satisfy the following constraints
\beq
\partial_1^n\Pop(x_1,0)=0=\partial_2^n\Pop(x_1,0),\quad n\ge1,
\label{grncons}
\eeq
which are derived from \rref{prchixnbc},\ts \rref{prchiynbc}\ts and
$g=1-2\Pop$ relation.
As in the case of the $O(N)$ and principal chiral sigma model
this would exclude any non-trivial solutions
with real analytic dependence on $x_2$.

\medskip

It is instructive to see that the simplest solutions of the ${\rm CP}^{N-1}$
and
Grassmannian sigma models, the (anti) instanton solutions (holomorphic maps)
do not satisfy the free boundary condition \rref{grfbc}, unless it is a trivial
(constant) solution.
Let us consider the $Gr(N,m)$ model. The general instanton solution
is given by \cite{WM}
\beq
X=F(F^\dagger F)^{-1/2},\quad \Pop=F(F^\dagger F)^{-1}F^\dagger,
\label{instsol}
\eeq
in which $F$ is an $N\times m$ matrix consisting of
linearly independent holomorphic vectors $f_1,\ldots,f_m$, which can be
chosen arbitrarily,
\beq
F=F(x_+)=\left(f_1(x_+),\ldots,f_m(x_+)\right),\quad \partial_-F=0.
\label{holomvec}
\eeq
Here  use is made of the complex two dimensional coordinates
$x_\pm=x_1\pm ix_2$.
It is easy to see
\beq
\partial_+\Pop=({\cal D}_+F)(F^\dagger F)^{-1}F^\dagger,\quad {\rm and}
\quad
\partial_-\Pop=F(F^\dagger F)^{-1}({\cal D}_+F)^\dagger,
\label{Pderi}
\eeq
where the `covariant derivative'  operator ${\cal D}_+$ is defined by
$$
{\cal D}_+F=\partial_+F-F(F^\dagger F)^{-1}F^\dagger\partial_+F
=(1-\Pop)\partial_+F.
$$
 From \rref{Pderi}\ts and the properties of the projection operator it
is obvious
\beq
\Pop(\partial_+\Pop)=0=(\partial_-\Pop)\Pop.
\label{insteq}
\eeq
By making a combination $\partial_-[\Pop(\partial_+\Pop)]-
\partial_+[(\partial_-\Pop)\Pop]$, we easily find that \Pop\ts satisfies
the equation of motion
$$
[\Pop, \partial_+\partial_-\Pop]=0.
$$
For more general solutions, see \cite{RSa}.

\medskip
Now let us consider the boundary condition of the instanton solution.
The free boundary condition \rref{grfbc}\ts is rewritten as
$$
i(\partial_+-\partial_-)\Pop(x_1,0)=0,
$$
or by using \rref{Pderi}
\beq
\left(({\cal D}_+F)(F^\dagger F)^{-1}F^\dagger-
F(F^\dagger F)^{-1}({\cal D}_+F)^\dagger\right)=0,
\label{insbc}
\eeq
It is easy to see $F^\dagger{\cal D}_+F=0$.
Multiplying $F$ from the right to \rref{insbc}, we obtain
\beq
({\cal D}_+F)(x_1,0)=0.
\label{insfbcsol}
\eeq
This in turn means
$$\partial_+\Pop(x_1,0)=0=\partial_-\Pop(x_1,0).
$$
Therefore  all the holomorphic vectors $f_1,\ldots,f_m$ are constants and
the instanton solution of the $Gr(N,m)$ model
satisfying the free boundary condition is trivial.
It should be noted that the conservation of the second charge is not
imposed here.

\section{Summary and Comments}
\setcounter{equation}{0}

We have shown that at the classical level the integrable non-linear sigma
models,
for example,  the $O(N)$,
the principal chiral, the ${\rm CP}^{N-1}$ and the complex Grassmannian
sigma models lose the integrability when restricted on a half plane
with free boundary.
Because of the conformal invariance of the equation of motion and the
Riemann mapping theorem, the result is valid in any simply connected domain
in two dimensional space with free boundary.
This result is in sharp contrast with other integrable models, sine-Gordon,
non-linear Schr\"odinger and affine Toda field theories
which are known to be integrable with various boundary conditions including the
free boundary.

However, the situation of the non-linear sigma models might be a rule rather
than an exception.
Let us consider a well known integrable field theory on the whole line: the KdV
equation
\beq
u_t=u_{xxx}-6uu_x,\quad u=u(t,x),\quad u_t=\partial_tu,
\quad u_x=\partial_xu,\quad{\rm etc.}
\label{kdveq}
\eeq
If we restrict it to a half line $0\leq x<\infty$, the first and second charges
$I^{(1)}$ and $I^{(2)}$
$$
I^{(1)}=\int_0^\infty u\ts dx,\quad I^{(2)}={1\over2}\int_0^\infty u^2 dx,
$$
are not conserved for either of the boundary conditions:
$$
u(t,0)=0,\quad {\rm or} \quad u_x(t,0)=0.
$$
It is easy to see
\Bears
{d\over{dt}}I^{(1)}(t)&=&-[u_{xx}-3u^2]\Bigm|_{x=0}\neq0,\\
{d\over{dt}}I^{(2)}(t)&=&-[uu_{xx}-{1\over2}u_x^2-2u^3]\Bigm|_{x=0}\neq0.
\Enars
It should be noted that the KdV equation does not have a
Lagrangian formulation.
It is easy to find some other integrable equations which lose integrability
when restricted to a half line.

\medskip
Here we would like to comment on the well known correspondence
(reduction) of
the $O(3)$ sigma model and the sine-Gordon theory on the full plane
established by Pohlmeyer \cite{Poh}.
This correspondence, if true on the  half plane, would lead to a
contradiction, since the $O(3)$ sigma model with a free boundary is
not integrable but,  as mentioned several times,
the sine-Gordon theory on the half plane is
integrable with the free and the other boundary interactions.
The correspondence is achieved by utilising the full conformal
invariance of the theory in the  two-dimensional Minkowski space;
\begin{equation}
	\xi\to \xi\p=f(\xi),\quad \eta\to\eta\p=g(\eta), \qquad \xi=(t+x)/2,\quad
	\eta=(t-x)/2,
	\label{conftr}
\end{equation}
where $f$ and $g$ are arbitrary (the  `left' and `right' transformation)
functions.
By appropriate choice of the  functions $f$ and $g$ one can transform the
   $O(3)$ sigma model variable $\phi$, such that it satisfies
\begin{equation}
	(\phi)^2=(\partial_\xi\phi)^2=(\partial_\eta\phi)^2=1.
	\label{redsys}
\end{equation}
Thus the only remaining degree of freedom is an  `angle' $\Psi$
\begin{equation}
	\cos\Psi=\partial_\xi\phi\cdot\partial_\eta\phi,
	\label{Psidef}
\end{equation}
which should satisfy the sine-Gordon equation
\begin{equation}
	\partial_\xi\partial_\eta\Psi+\sin\Psi=0.
	\label{sgeq}
\end{equation}

On the half plane, on the other hand, the conformal
transformation depends on  only one arbitrary function $h$,
\begin{equation}
	\xi\p=h(\xi),\quad \eta\p=h(\eta),
	\label{hftr}
\end{equation}
which is not sufficient to reduce the system to the form \rref{redsys}.
Thus the $O(3)$ sigma model cannot be reduced to the sine-Gordon theory
on the half plane.
\medskip

It is very challenging to consider the {\em quantum integrability} of
non-linear sigma models on the half plane.
As mentioned in section 1  the Reflection equation
of these models offers interesting extensions of the Yang-Baxter equations
and its related algebras for the known exact factorisable S-matrices.
It should be remarked that the relationship between the classical and quantum
theories of non-linear sigma models is not so straightforward as that of
sine-Gordon and/or affine Toda field theories.
At the classical level the non-linear sigma models are conformally invariant,
i.e., massless, whereas the quantum spectrum consists of massive particles
belonging to
certain representations of appropriate Lie algebras.

\section*{Acknowledgments}
We thank E.\ts Corrigan, K.\ts Fujii, R.\ts Kobayashi, N.\ts Otsuki,
M.\ts Oka, A.\ts Yoshioka, C.\ts Zachos and W.\ts Zakrzewski
 for interesting  comments and discussion. We thank A.\ts Yoshioka for
 bringing \cite{Ham}\ts into our attention.


\def\CMP{{\it Comm.\ts Math.\ts Phys.\ts}}
\def\FAP{{\it Funct.\ts Analy.\ts Appl.\ts}}
\def\IJMP{{\it Int.\ts J.\ts Mod.\ts Phys.\ts}}
\def\JMP{{\it J.\ts Math.\ts Phys.\ts}}
\def\JP{{\it J.\ts Phys.\ts}}
\def\LMP{{\it Lett.\ts Math.\ts Phys.\ts}}
\def\MPL{{\it Mod.\ts Phys.\ts Lett.\ts}}
\def\NP{{\it Nucl.\ts Phys.\ts}}
\def\PL{{\it Phys.\ts Lett.\ts}}
\def\PR{{\it Phys.\ts Rev.\ts}}
\def\PRL{{\it Phys.\ts Rev.\ts Lett.\ts}}
\def\PTP{{\it Prog.\ts Theor.\ts Phys.\ts}}
\def\TAMS{{\it Trans.\ts Amer.\ts Math.\ts Soc.\ts}}
\def\TMP{{\it Theor.\ts Math.\ts Phys.\ts}}
\def\Zm{Zamolodchikov}
\def\ZP{{\it Zeit.\ts f.\ts Phys.\ts}}
\def\AZm{A.\ts B.\ts \Zm}
\def\dur{H.\ts W.\ts Braden, E.\ts Corrigan, P.\ts E.\ts Dorey and R.\ts
Sasaki}

\end{document}